\documentclass[aps,prl,twocolumn,showpacs,superscriptaddress]{revtex4-2} 

\usepackage{graphicx}
\usepackage{xr-hyper}  
\usepackage[colorlinks,citecolor=blue,urlcolor=blue,linkcolor=blue]{hyperref}
\usepackage{siunitx}	
\usepackage{amsmath}
\usepackage{amsthm}
\usepackage{amssymb}
\usepackage{amsthm}
\usepackage{braket}
\usepackage{booktabs}
\usepackage{footmisc}
\usepackage{mathrsfs}
\usepackage{cleveref}
\usepackage{dsfont}
\usepackage{subfigure}
\usepackage{multirow}
\usepackage{enumerate}
\usepackage{lineno}

\makeatletter
\def\maketitle{
\@author@finish
\title@column\titleblock@produce
\suppressfloats[t]}
\makeatother

\newcommand{\bracket}[2]{\ensuremath{\langle#1 \vphantom{#2}| #2\vphantom{#1}\rangle}}

\newcommand{\ketbra}[2]{\ensuremath{|#1 \vphantom{#2}\rangle \langle #2\vphantom{#1}|}}

\newcommand{\sgn}{\mathrm{sgn}}

\newcommand{\tr}[1]{\mathrm{Tr}\left(#1\right)}

\newcommand{\trpp}[2]{\mathrm{Tr}_{#1}\left(#2\right)}
\newcommand{\abs}[1]{\left| #1 \right|} 

\newcommand{\kb}{k_\mathrm{B}}

\newcommand{\im}{\mathrm{i}}

\newcommand{\sg}[1]{\mathrm{S}_{#1}}

\begin{document}
\title{Distinguishability-induced many-body decoherence}

\author{Christoph Dittel}
\email{christoph.dittel@physik.uni-freiburg.de}
\affiliation{Physikalisches Institut, Albert-Ludwigs-Universit{\"a}t Freiburg, Hermann-Herder-Straße 3, 79104 Freiburg, Germany}
\affiliation{EUCOR Centre for Quantum Science and Quantum Computing, Albert-Ludwigs-Universität Freiburg, Hermann-Herder-Straße 3, 79104 Freiburg, Germany}
\affiliation{Freiburg Institute for Advanced Studies, Albert-Ludwigs-Universität Freiburg, Albertstraße 19, 79104 Freiburg, Germany}
\affiliation{Department of Physics, Lund University, Box 118, 221 00 Lund, Sweden}

\author{Andreas Buchleitner}
\affiliation{Physikalisches Institut, Albert-Ludwigs-Universit{\"a}t Freiburg, Hermann-Herder-Straße 3, 79104 Freiburg, Germany}
\affiliation{EUCOR Centre for Quantum Science and Quantum Computing, Albert-Ludwigs-Universität Freiburg, Hermann-Herder-Straße 3, 79104 Freiburg, Germany}

\date{\today}

\begin{abstract}
We show that many-body interference (MBI) phenomena are exponentially suppressed in the particle number, if only the identical quantum objects brought to interference acquire a finite level of distinguishability through statistical mixing of some internal, unobserved degrees of freedom. We discuss consequences for cold atom and photonic circuitry experiments.
\end{abstract}

\maketitle

Wave-particle duality, the modulation of the statistics of particle-like detection events by wave-like interference patterns, is the essential feature which distinguishes the quantum from the classical realm. It hinges on a sufficient degree of purity of the interfering quantum object's state \cite{Ghirardi-UD-1986,Brune-MP-1992,Brune-OP-1996,Nakamura-CC-1999,Friedman-QS-2000,Zurek-DE-2003,Hackermueller-DM-2004,Schlosshauer-DM-2005,Hornberger-QI-2012,Nimmrichter-MM-2013,Arndt-TL-2014,Schlosshauer-QD-2019,Delic-CL-2020,Dittel-WP-2021}, as the precondition to witness interference in a suitably chosen measurement set-up which probes the quantum  state's coherences in the associated basis. On the many-body level, coherent superpositions of many-body states give rise to multi-partite entanglement \cite{Mintert-MD-2005,Tichy-EI-2013,Ketterer-CM-2019,Benatti-EI-2020}, and to many-body interference (MBI) phenomena \cite{Hong-MS-1987,Lim-ME-2005,Tichy-ZT-2010,Tichy-MP-2012,Aaronson-CC-2013,Spring-BS-2013,Crespi-IM-2013,Tillmann-EB-2013,Shchesnovich-PI-2015,Menssen-DM-2017,Flamini-PQ-2019,Tichy-PhDThesis-2011,Walschaers-PhDThesis-2016,Bruenner-PhDThesis-2018,Dittel-TD-2018,Dittel-AI-2019,Njoya-PhdThesis2022,Brunner-PhDThesis-2023,Seron-PhdThesis-2023,Englbrecht-PhDThesis-2023} when the involved elementary constituents are identical and at least partially indistinguishable. Entanglement (i.e., non-separability) between those degrees of freedom (dof) which are interrogated by the experimental measurement set-up and other, unobserved (``environmental", ``bath",  ``internal" or ``ancilla") dof -- which are traced over when sampling the measurement record -- reduces the purity of the quantum object's state in its observed dof, and thus its ability to exhibit interference phenomena, by reducing the strength of the associated coherences \cite{Mayer-CS-2011,Walschaers-FM-2016,Dittel-WP-2021,Brunner-MB-2023}. It is intuitively plausible that the larger the number of involved dof and of constituents, it becomes ever more difficult to warrant separability of observed and unobserved dof (by preventing the former from any type of interactions with the latter). This is the fundamental impediment to push the quantum-classical demarcation line to meso- if not macroscopic scales. 

Notwithstanding, stunning progress has been achieved in preparing coherent superposition states of collective degrees of freedom of many-body compounds of ever increasing size, from bucky balls \cite{Arndt-WP-1999,Hackermueller-DM-2004,Hornberger-QI-2012} to supercurrents \cite{Nakamura-CC-1999,Nimmrichter-MM-2013}, micromechanical oscillators \cite{Teufel-SC-2011,Chan-LC-2011}, and Bose Einstein Condensates \cite{Andrews-OI-1997,Wallis-IT-1998}. Furthermore, beyond such experiments, which probe effective single-body coherences, experimental progress in the manipulation of controlled many-body quantum systems on the level of single constituents \cite{Bakr-QG-2009,Sherson-SA-2010,Bayha-OE-2020,Crespi-IM-2013,Spring-BS-2013,Tillmann-EB-2013,Meinert-OM-2014,Preiss-SC-2015,Roos-RQ-2017,Menssen-DM-2017,Flamini-PQ-2019,Muenzberg-SA-2021} now allows to assess bona fide MBI phenomena. While on the single-body level the quantum object’s effective size or mass defines the scale on which interference phenomena are to be observed \cite{Nimmrichter-MM-2013}, it is suggestive that the relevant scale is defined by the number $N$ of interfering constituents on the MBI level. Our present purpose is to make this quantitative. We derive a scaling law which shows that distinguishability due to a finite degree of mixedness in the quantum objects' internal, ancilla degrees of freedom -- easily brought about by some residual environment coupling -- induces the exponential suppression of many-body coherences and hence of MBI phenomena with increasing $N$. We examine and discuss the consequences of our scaling law for MBI of cold atoms or of photons in optical lattices or in photonic circuits, respectively.

Consider a quantum many-body system composed of $N$ identical bosons or fermions localized in mutually orthogonal external states (e.g., think of atoms in a Mott state \cite{Bakr-QG-2009,Sherson-SA-2010}, or of photons in distinct optical modes \cite{Spring-BS-2013,Crespi-IM-2013}). To describe the particles' distribution across their individual external states, we make use of the first quantization formalism and denote the $N$-particle basis states by $\ket{\vec{E}}=\ket{E_1}\otimes \cdots \otimes \ket{E_N}$, where $\ket{E_\alpha}$ is the external state occupied by the $\alpha$th particle, with $\bracket{E_\alpha}{E_\alpha'}=\delta_{E_\alpha,E_\alpha'}$ and $\bracket{\vec{E}}{\vec{E}'}=\prod_{\alpha=1}^N \delta_{E_\alpha,E'_\alpha}$ \cite{Dittel-WP-2021}. Further suppose that the particles are equipped with internal degrees of freedom (e.g., the arrival time and polarization state of photons, or the atoms' electronic energy levels), prepared in potentially mixed internal many-body states $\rho$, which are neither acted upon nor measured.

Given $N$ identical bosons (fermions), we must (anti-) symmetrize $\ketbra{\vec{E}}{\vec{E}}\otimes \rho$ with respect to all particle permutations of the symmetric group $\mathrm{S}_N$ of $N$ elements. Since MBI is to be observed by interrogation of the external degrees of freedom alone, we subsequently trace over the internal degrees of freedom. This yields the external many-body state $\rho_\mathrm{E}= \sum_{\pi,\pi' \in \sg{N}} [\rho_\mathrm{E} ]_{\pi,\pi'} \ketbra{\vec{E}_\pi}{\vec{E}_{\pi'}}$,
\begin{align}\label{eq:rhoEElements}
[\rho_\mathrm{E} ]_{\pi,\pi'}=(-1)_\mathrm{B(F)}^{\pi\pi'} \frac{1}{N!} \tr{\Pi_\pi \rho \Pi^\dagger_{\pi'}}\, ,
\end{align}
which has $N!\times N!$ matrix elements \cite{Dittel-AI-2019,Dittel-WP-2021,Minke-CF-2021}. The many-particle basis states $\ket{\vec{E}_\pi}=\ket{E_{\pi(1)}}\otimes \dots \otimes \ket{E_{\pi(N)}}$ of $\rho_\mathrm{E}$ result from $\ket{\vec{E}}$ by permuting the particles according to $\pi^{-1} \in \sg{N}$, much as the operator $\Pi_\pi$ in~\eqref{eq:rhoEElements} performs a particle permutation $\pi^{-1}$ in the internal degrees of freedom. $(-1)^{\pi\pi'}_\mathrm{B}=1$ for bosons, with $\pi\pi'$ a composition of $\pi$ and $\pi'$, and $(-1)^{\pi\pi'}_\mathrm{F}=\sgn(\pi\pi')$ for fermions.

Many-body coherences $[\rho_\mathrm{E} ]_{\pi,\pi'}$, $\pi\neq\pi'$ in~\eqref{eq:rhoEElements}, thus result from the (anti-) symmetrization and are associated with different orderings of the particles in the tensor product structure. By virtue of the trace in~\eqref{eq:rhoEElements}, these coherences are governed by the particles' mutual indistinguishability \cite{Shchesnovich-PI-2015,Dittel-WP-2021,Minke-CF-2021} with respect to their internal degrees of freedom. In the limiting case of pure states of separable, perfectly indistinguishable bosons ($\mathrm{B}$) or fermions ($\mathrm{F}$), i.e., $\rho=\ketbra{\phi}{\phi}$ with $\ket{\phi}=\ket{\varphi}\otimes \dots \otimes \ket{\varphi}$, the trace in~\eqref{eq:rhoEElements} yields unity for all $\pi,\pi'\in\sg{N}$, such that the corresponding reduced external state is fully coherent and described by a pure state, $\rho_\mathrm{E}^\mathrm{B(F)}=\ketbra{\psi_\mathrm{B(F)}}{\psi_\mathrm{B(F)}}$, with $\ket{\psi_\mathrm{B(F)}}=\sum_{\pi\in\sg{N}}(-1)_\mathrm{B(F)}^{\pi} \ket{\vec{E}_\pi}/\sqrt{N!}$ the usual Fock state of indistinguishable particles. On the other hand, pure states of separable, fully distinguishable ($\mathrm{D}$) particles feature internal states with orthogonal support, i.e., $\rho=\ketbra{\phi}{\phi}$ with $\ket{\phi}=\ket{\varphi_1}\otimes \dots \otimes \ket{\varphi_N}$ and $\bracket{\varphi_j}{\varphi_k}=\delta_{j,k}$, and thus give rise to a fully incoherent many-body state, $\rho_\mathrm{E}^\mathrm{D}=\sum_{\pi\in\sg{N}}\ketbra{\vec{E}_\pi}{\vec{E}_\pi}/N!$. The normalized many-body coherence \cite{Dittel-AI-2019,Dittel-WP-2021} of $\rho_\mathrm{E}$ is thus given by
\begin{align}\label{eq:WC}
\mathcal{W}_\mathrm{C}&= \frac{1}{N!-1} \sum_{\substack{\pi,\pi' \in \sg{N} \\ \pi \neq \pi'}} \abs{ [\rho_\mathrm{E} ]_{\pi,\pi'}}\, ,\, 0\leq \mathcal{W}_\mathrm{C} \leq 1\, ,
\end{align}
with pure states of separable, fully distinguishable (indistinguishable) particles saturating the lower (upper) bound. Since the modulus in Eq.~\eqref{eq:WC} erases the sign of the coherences~\eqref{eq:rhoEElements}, the results presented hereafter apply for bosons as well as for fermions. 

Another source of distinguishability is due to mixedness of the particles' internal degrees of freedom, which typically arises through decoherence on the single-particle level \cite{Zurek-DE-2003,Schlosshauer-DM-2005,Schlosshauer-QD-2019}: Suppose that each particle is well described by the same single-particle state $\rho_\mathrm{1p}$ \cite{SC-Shchesnovich-2014,Marshall-DI-2022}, such that $\rho=\rho_\mathrm{1p}\otimes \cdots \otimes \rho_\mathrm{1p}$ in Eq.~\eqref{eq:rhoEElements}. In Sec.~II of \footnote{Supplemental Material} we show that in this case the normalized coherence $\mathcal{W}_\mathrm{C}$ of many  bosons (fermions), $N!\gg 1$, can be identified with the expectation value of the projector $\Pi_\mathrm{S(A)}$ onto the $N$-particle (anti-) symmetric subspace, $\mathcal{W}_\mathrm{C}\approx\tr{\Pi_\mathrm{S(A)}\rho_\mathrm{E}}$ 
\footnote{Note that for a bosonic state $\rho_\mathrm{E}$, tight lower bounds of $\tr{\Pi_\mathrm{S}\rho_\mathrm{E}}$ can be efficiently measured \cite{Englbrecht-II-2024}.}, which, as we further show, is equivalent to the support of the unsymmetrized $N$-particle internal state $\rho$ on the $N$-particle symmetric subspace, i.e., $\mathcal{W}_\mathrm{C}\approx\tr{\Pi_\mathrm{S(A)}\rho_\mathrm{E}}=\tr{\Pi_\mathrm{S}\rho}$. Although the internal states of all particles are described by the same density operator, the particles are indistinguishable if and only if $\rho_\mathrm{1p}$ is pure. To make this explicit, suppose that $\rho_\mathrm{1p}$ has a discrete spectrum of $m$ eigenvalues $\lambda_j\geq 0$ with corresponding eigenvectors $\ket{j}$, such that its eigendecomposition reads 
$\rho_\mathrm{1p}=\sum_{j=1}^m \lambda_j \ketbra{j}{j}$. As we show in Sec.~III of \cite{Note1}, the normalized coherence~\eqref{eq:WC} can then be written, for $N!\gg 1$, as
\begin{align}\label{eq:WCeigen}
\mathcal{W}_\mathrm{C}&\approx\sum_{J_1+J_2+\cdots+J_m=N} \lambda_1^{J_1}\lambda_2^{J_2} \cdots \lambda_m^{J_m},
\end{align}
with the sum running over all non-negative integers $J_1,J_2,\dots,J_m$ summing to $N$. Equation~\eqref{eq:WCeigen} explicitly shows how the spectrum of $\rho_\mathrm{1p}$ controls many-body coherence, with $\mathcal{W}_\mathrm{C}= 1$ for pure $\rho_\mathrm{1p}$, and 
\begin{align}\label{eq:WCDistLim}
\mathcal{W}_\mathrm{C} \approx  \frac{1}{m^N} {N+m-1 \choose m-1}\, ,
\end{align}
for $\rho_\mathrm{1p}$ maximally mixed. Since $\mathcal{W}_\mathrm{C} \approx\tr{\Pi_\mathrm{S}\rho}$, the finite residual coherence quantified by (\ref{eq:WCDistLim}) is given by the relative dimension of the symmetric component $\mathrm{S_N}(\mathcal{H}_\mathrm{I})$ with respect to that of the total internal Hilbert space $\mathcal{H}_\mathrm{I}$. In the limit of an infinite number $m$ of single-particle internal states, $\mathrm{S_N}(\mathcal{H}_\mathrm{I})$ tends to zero, and so does the residual coherence in~\eqref{eq:WCDistLim}.

With $\mathcal{W}_\mathrm{C}$ from Eq.~\eqref{eq:WC} at hand, we have a quantifier of many-body coherence, i.e., a quantifier of the very source of MBI, independently of the exact experimental protocol \cite{Dittel-AI-2019,Dittel-WP-2021}. Any such experiment, however, can only unfold the complexity seeded by MBI for large system sizes. We therefore focus on the scaling behavior of $\mathcal{W}_\mathrm{C}$ in the thermodynamic limit, $N\rightarrow \infty$ (at fixed particle density $N/L$, with $L$ the dimension of the external single-particle Hilbert space, i.e., here, the number of external single-particle modes), and further examine the scaling behavior in the special case of faint particle distinguishability, i.e., for very weakly mixed internal states $\rho_\mathrm{1p}$. 
\begin{figure*}[t]
\centering
\includegraphics[width=\linewidth]{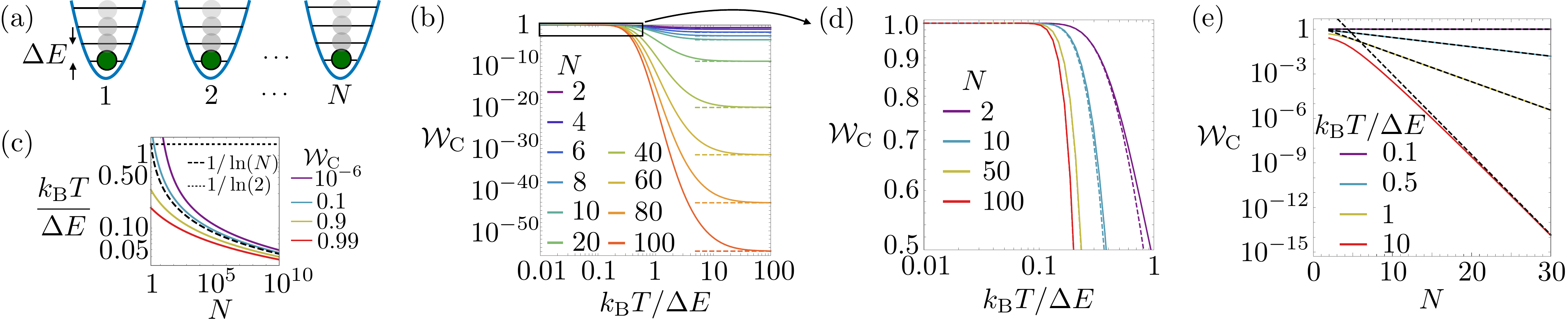}
\caption{Many-body decoherence induced by the thermal population of energy levels of (ultra-) cold atoms in individual optical lattice sites. (a) $N$ atoms (green filled circles) occupy one lattice site (blue parabolic wells, representing the atoms' external states) each, and exhibit a Boltzmann population distribution (gray blurring) at temperature $T$ over the (internal) states, modeled by four energy levels (with energy difference $\Delta E=E_{j+1}-E_j$). (b) Normalized many-body coherence $\mathcal{W}_\mathrm{C}$ vs. temperature $\kb T/\Delta E$, for $N=2\ldots 100$. Dashed lines represent the infinite temperature limit \eqref{eq:WCDistLim}. (c) Admissible thermal excitation $\kb T/\Delta E$ given by \eqref{eq:kbTvsN}, as a function of $N$, at different target coherences $\mathcal{W}_\mathrm{C}$. (d) Zoom into (b), together with a comparison to the approximation $\mathcal{W}_\mathrm{C}\approx (1-e^{-\beta\Delta E})^N$ (dashed lines), valid in the limit $\kb T/\Delta E \ll 1/\ln(2)\simeq 1.44$ and $N!\gg 1$, for $N=2,10,50,100$. (e) $\mathcal{W}_\mathrm{C}$ as a function of the particle number $N$ (solid lines), for different temperatures $\kb T/\Delta E$. Black dashed lines indicate the exponential asymptotics \eqref{eq:WCThLimit} (with the eigenvalues of the four dimensional thermal state obtained by diagonalisation) in the thermodynamic limit $N\rightarrow \infty$, $N/L={\rm const.}$}
\label{fig:Gibbs}
\end{figure*}

Let $\lambda_\mathrm{max}=\max_{j}\lambda_j$ be the maximum eigenvalue of $\rho_\mathrm{1p}$, and suppose that $\lambda_\mathrm{max}$ is non-degenerate. Unless all particles are perfectly indistinguishable (i.e., $\lambda_\mathrm{max}=1$), it then follows from Eq.~\eqref{eq:WCeigen} (see Sec.~IV of \cite{Note1}) that, in the thermodynamic limit, 
\begin{align}\label{eq:WCThLimit}
\mathcal{W}_\mathrm{C} \approx \lambda_\mathrm{max}^N \prod_{\substack{j=1\\ j\neq \mathrm{max}}}^{m} \left( 1-\frac{\lambda_j}{\lambda_\mathrm{max}}\right)^{-1}\, , 
\end{align}
i.e., $\mathcal{W}_\mathrm{C}$ from Eq.~\eqref{eq:WCeigen} vanishes exponentially in the number $N$ of constituents \footnote{$d$-fold degeneracy of $\lambda_\mathrm{max}$ instead leads to $\mathcal{W}_\mathrm{C} \propto (N+1)^{d-1} \lambda_\mathrm{max}^N$ (Sec.~IV of \cite{Note1})}. Alternatively, we decompose the internal single-particle state into a dominating pure and a faint mixed component, $\rho_\mathrm{1p}=(1-\epsilon)\ketbra{\phi}{\phi}+\epsilon \tilde{\rho}_\mathrm{1p}$, with $\epsilon \ll 1/2$, and $\tilde{\rho}_\mathrm{1p}$ a valid density operator. Again (Sec.~V in \cite{Note1}), for $N!\gg 1$, $\mathcal{W}_\mathrm{C}$ exhibits exponential scaling: 
\begin{align}\label{eq:WCdecay}
\mathcal{W}_\mathrm{C}\approx (1-\epsilon)^N\, .
\end{align}

To asses the scaling behavior of the entire hierarchy from two- to N-body coherences, we consider the reduced $k$-particle state $\rho_\mathrm{E}^{(k)}=\trpp{N-k}{\rho_\mathrm{E}}$ \cite{Minke-CF-2021,Brunner-MC-2022,Brunner-PhDThesis-2023} obtained from $\rho_\mathrm{E}$ by tracing out all but $k$ particles. As compared to $\rho_\mathrm{E}$, the reduced $k$-particle state $\rho_\mathrm{E}^{(k)}$ carries no information about collective many-body properties of subsets of more than $k$ particles \cite{Brunner-MC-2022}. The normalized coherence $\mathcal{W}_\mathrm{C}^{(k)}$ of $\rho_\mathrm{E}^{(k)}$ (see Sec.~VI of \cite{Note1}) thus quantifies $k$-particle coherence of order $k<N$. Since we consider the particles to be localized in distinct external states, with equal, independent internal states $\rho_\mathrm{1p}$, it is intuitively clear that, for $k!\gg 1$, $\mathcal{W}_\mathrm{C}^{(k)}$ takes a similar form as Eq.~\eqref{eq:WCeigen} (see Sec.~VI of \cite{Note1}): $\mathcal{W}_\mathrm{C}^{(k)}\approx\sum_{J_1+J_2+\cdots+J_m=k} \lambda_1^{J_1}\lambda_2^{J_2} \cdots \lambda_m^{J_m}$. Consequently, Eqs.~\eqref{eq:WCThLimit} and~\eqref{eq:WCdecay} also apply to $\mathcal{W}_\mathrm{C}^{(k)}$, but with an exponential scaling in $k$ instead of $N$, such that $k(<N)$-body coherences fade away slower than those of order $N$. Note that $\mathcal{W}_\mathrm{C}^{(k)}$ is, by the very purpose of its construction, undefined for $k=1$, and that the constituent particles interference with themselves is controlled by the coherence of the reduced single-particle state, as in standard single-body interference.

We now discuss the consequences of the above for specific experimental scenarios. First consider cold atoms in distinct optical lattice sites (external states) \cite{Bakr-QG-2009,Sherson-SA-2010,Meinert-OM-2014,Preiss-SC-2015} which we model as harmonic potentials each with $m=4$ equidistant energy levels (internal states) with energy differences $E_{j+1}-E_j=\Delta E$ [see Fig~\ref{fig:Gibbs}(a)]. The atoms' population distribution over the oscillator levels be given by a Boltzmann distribution at equilibrium temperature $T$ [see Fig.~\ref{fig:Gibbs}(a)]. Their single-particle internal states $\rho_\mathrm{1p}$ are then given as $\rho_\mathrm{1p}= \sum_{j=1}^{m} e^{-\beta E_j}Z(\beta)^{-1} \ketbra{j}{j}$, with $\beta=1/\kb T$ the inverse temperature, $\kb$ the Boltzmann constant, $Z(\beta)=\sum_{j=1}^{m} e^{-\beta E_j}$ the partition function, and $\{\ket{j}\}_{j=1}^{m}$ a set of $m$ orthonormal oscillator energy states, with associated eigenvalues $\lambda_j=e^{-\beta E_j} Z(\beta)^{-1}$. For particle numbers up to $N=100$, Figs.~\ref{fig:Gibbs}(b,d) show a monotonous decrease of $\mathcal{W}_\mathrm{C}$ with increasing $T$, from a fully coherent [$k_BT/\Delta E\lessapprox 0.1$, see Fig.~\ref{fig:Gibbs}(c)] to an almost incoherent ($k_BT/\Delta E\gtrapprox 1$) external many-body state with finite residual coherence as described by Eq.~\eqref{eq:WCDistLim}. The complementary, asymptotically exponential decrease of $\mathcal{W}_\mathrm{C}$ as a function of the particle number $N$, at fixed temperature, is 
shown in Fig.~\ref{fig:Gibbs}(e) -- in perfect agreement with \eqref{eq:WCThLimit} [dashed lines in Fig.~\ref{fig:Gibbs}(e)].

Close to the critical temperature, $0.1\lessapprox k_BT/\Delta\lessapprox 1$, at which external decoherence sets in in Fig.~\ref{fig:Gibbs}(d), we can apply Eq.~\eqref{eq:WCdecay}: With $e^{-\beta\Delta E} \ll 1/2$, i.e., $\kb T /\Delta E\ll 1/\ln(2)$, the internal ground state (i.e., lowest energy level of the local oscillator potential) dominates, leading to (see Sec.~VII of \cite{Note1}) $\mathcal{W}_\mathrm{C}\approx (1-e^{-\beta\Delta E})^N$ [dashed lines in Fig.~\ref{fig:Gibbs}(d)], such that 
\begin{align}\label{eq:kbTvsN}
\frac{\kb T}{\Delta E} \approx \frac{-1}{\ln\left(1-\mathcal{W}_\mathrm{C}^{1/N} \right)}\, .
\end{align}
If $e^{-\beta\Delta E} \ll 1/N$, i.e., $\kb T/\Delta E \ll 1/\ln(N)$, further approximation yields $\mathcal{W}_\mathrm{C}\approx 1-N e^{-\beta\Delta E}$ and $\kb T/\Delta E \approx 1/\ln[N/(1- \mathcal{W}_\mathrm{C})]$. For our example illustrated in Fig.~\ref{fig:Gibbs}(a), Eq.~\eqref{eq:kbTvsN} is plotted  in Fig.~\ref{fig:Gibbs}(c), for fixed 
coherences $\mathcal{W}_\mathrm{C}$, as a function of $N$. We see that $\kb T/\Delta E$ only gradually decreases for increasing $N$. This explains why the onset of decoherence, manifest in the drop of $\mathcal{W}_\mathrm{C}$ in Figs.~\ref{fig:Gibbs}(b,d), only marginally shifts towards smaller critical temperatures with increasing $N$. In particular, this observation suggests that MBI remains observable, with high visibilities, in experiments with large particle numbers, at temperatures $\kb T$ not much below $\Delta E$.


As a second example, let us assess the case of $N$ photons propagating along distinct, possibly coupled optical modes (external degrees of freedom) \cite{Spring-BS-2013,Crespi-IM-2013,Tillmann-EB-2013,Flamini-PQ-2019}. Different injection times (internal degrees of freedom) -- a typical error source in photonic experiments -- render the photons mutually partially distinguishable \cite{Hong-MS-1987,Ra-NQ-2013,Muenzberg-WP-2019}. As above, we describe the internal state of every photon by the same mixed single-particle internal state $\rho_\mathrm{1p}= \int_{-\infty}^\infty \mathrm{d}t \ P(t) \ketbra{t}{t}$, with arrival time probability distribution $P(t)$, and $\ket{t}=(2 \pi \Delta^2)^{-1/4} \int_{-\infty}^\infty \mathrm{d}\omega \ e^{\im \omega t} e^{-\frac{(\omega-\Omega)^2}{4\Delta^2}} \ket{\omega}$ a single photon's (internal) state with arrival time $t$, Gaussian frequency spectrum of spectral width $\Delta$ around the central frequency $\Omega$, and $\bracket{\omega}{\omega'}=\delta(\omega-\omega')$.
\begin{figure}[t]
\centering
\includegraphics[width=\linewidth]{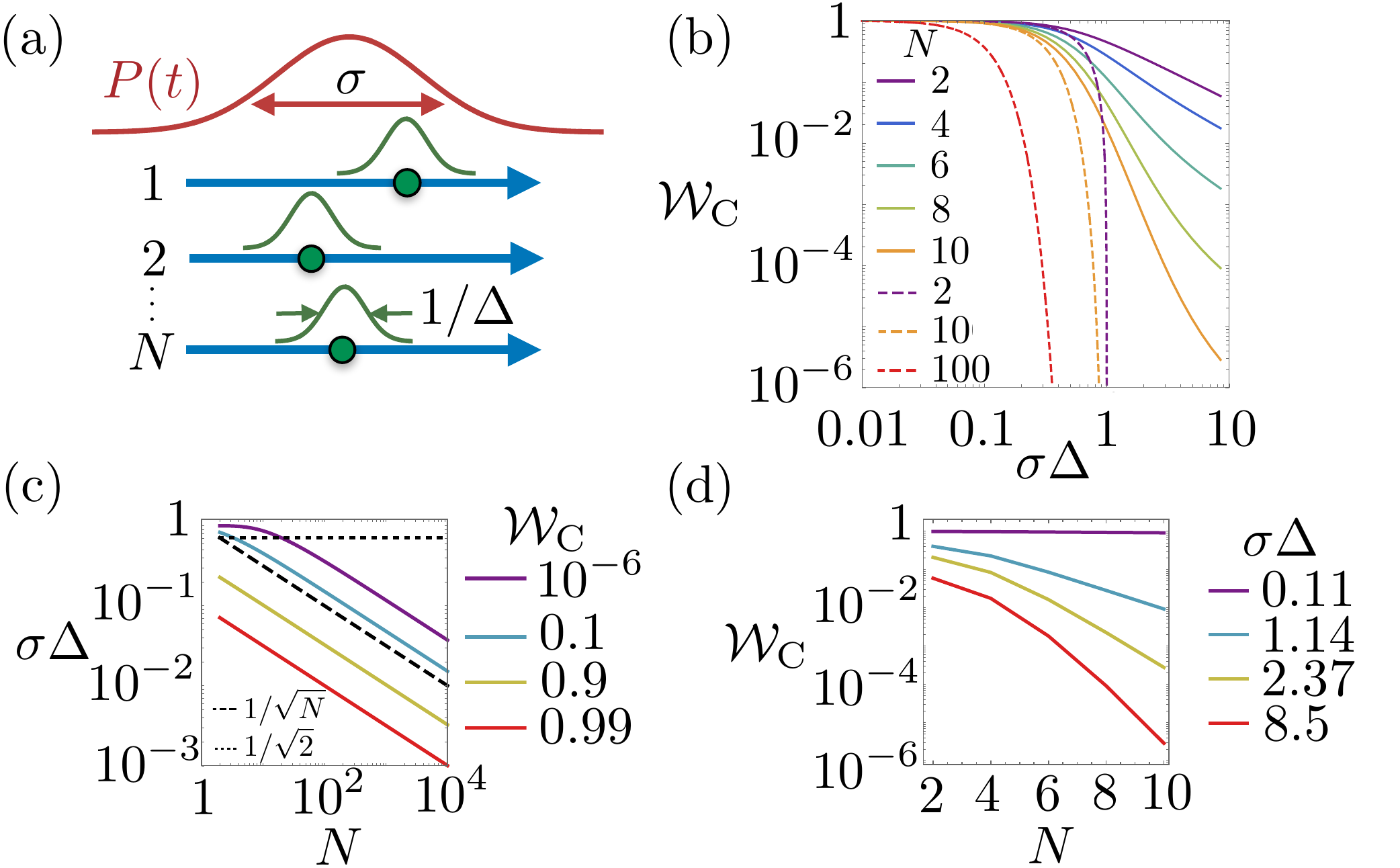}
\caption{Many-body decoherence induced by random arrival times of photons in optical modes. (a) $N$ photons (green filled circles) in distinct optical modes (blue arrows, representing the photons' external states) each have a Gaussian frequency spectrum with spectral width $\Delta$. Their random arrival times (green envelope, internal states) have a Gaussian probability distribution $P(t)$ (red envelope), with standard deviation $\sigma$. (b) Normalised many-body coherence $\mathcal{W}_\mathrm{C}$ (solid lines) vs. $\sigma\Delta$, for different particle numbers $N=2\ldots 10$. Dashed lines indicate the approximation $\mathcal{W}_\mathrm{C}\approx (1-\Delta^2\sigma^2)^N$, valid in the limit $\sigma\Delta\ll 1/\sqrt{2}$ and $N!\gg1$, for $N=2,10,100$. (c) Acceptable scatter $\sigma\Delta$ of the photons' arrival time given a desired coherence level $\mathcal{W}_\mathrm{C}$, as a function of $N$, in the limit $\sigma\Delta\ll 1/\sqrt{2}\simeq 0.7$ and $N!\gg1$ [see Eq.~\eqref{eq:DeltaSigma}]. For $\sigma\Delta \ll 1/\sqrt{N}$ (black dashed line), $\sigma\Delta \approx \sqrt{(1-\mathcal{W}_\mathrm{C})/N}$. (d) $\mathcal{W}_\mathrm{C}$ as a function of $N$ for various values of $\sigma\Delta$, indicating the convergence into an exponential decay in the limit of large $N$.}
\label{fig:Random}
\end{figure}
Figure~\ref{fig:Random} provides an example of normally distributed arrival times, $P(t)=\exp(-(t-t_0)^2/(2\sigma^2))/\sqrt{2\pi\sigma^2}$, with mean $\braket{t}=t_0$ and standard deviation $\sigma=(\braket{t^2}-\braket{t}^2)^{1/2}$, for which we calculate the behavior of $\mathcal{W}_\mathrm{C}$ as a function of $\sigma\Delta$ and $N$ via Eqs.~\eqref{eq:rhoEElements} and~\eqref{eq:WC} in the following.

The finite width of the photonic arrival time distribution in terms of the temporal width of a photonic wave packet, $\sigma\Delta$, now reduces, similar to the finite temperature distribution over atomic energy bands above, the normalized coherence $\mathcal{W}_\mathrm{C}$ of the $N$-photon state, plotted in Fig.~\ref{fig:Random}(b). We observe a qualitatively similar transition  from a fully coherent to a fully incoherent many-body state, as a function of $\sigma\Delta$. However, since the here considered internal degrees of freedom -- the arrival times -- are continuous, $\mathcal{W}_\mathrm{C}$ truly vanishes in the limit of large $\sigma\Delta$, since $m\rightarrow \infty$ in Eq.~\eqref{eq:WCDistLim}. While a direct application of Eq.~\eqref{eq:WCThLimit} requires the spectral decomposition of $\rho_\mathrm{1p}$, we here restrict to a qualitative observation, with Fig.~\ref{fig:Random}(d) indicating a convergence into an exponential decay of $\mathcal{W}_\mathrm{C}$ in the limit of large $N$. For $N!\gg 1$ and small distinguishabilities $\sigma\Delta \ll 1/\sqrt{2}$, we show in Sec.~VIII of \cite{Note1} that Eq.~\eqref{eq:WCdecay} can be reformulated as $\mathcal{W}_\mathrm{C} \approx (1-\sigma^2\Delta^2)^N$ (dashed lines in Fig.~\ref{fig:Random}(b); compare to the average mutual fidelity in \cite{SC-Shchesnovich-2014}). Accordingly, we have
\begin{align}\label{eq:DeltaSigma}
\sigma\Delta  \approx \sqrt{1-\mathcal{W}_\mathrm{C}^{1/N}}.
\end{align}
For very faint distinguishabilities $\sigma\Delta \ll 1/\sqrt{N}$, this simplifies to $\sigma\Delta \approx \sqrt{(1-\mathcal{W}_\mathrm{C})/N}$. In Fig.~\ref{fig:Random}(c) we plot Eq.~\eqref{eq:DeltaSigma} for our example of normally distributed arrival times. In particular, for $\sigma\Delta \ll 1/\sqrt{N}$, it confirms the power law $\sigma \Delta \propto N^{-1/2}$ for a targeted level of coherence $\mathcal{W}_\mathrm{C}$, with $\sigma\Delta $ sharply decreasing with increasing particle number $N$. Thus, the photons' arrival times must be increasingly well controlled (with respect to their inverse spectral width) in order to harvest the interference of an increasing number of particles.

We have thus quantified the challenge to witness mutual interference of a large number of identical particles, given that, the larger this number, the more difficult to prevent the system constituents from interactions with environmental degrees of freedom. Our present analysis relies on the ``static" description of a given many-body state with finite mixedness of its individual constituents, and thereby sets limits to the acceptable level of noise if a certain level of many-body coherence is to be guaranteed. What is not resolved in this analysis is the time dependence of the decoherence process, and a dynamical description of many-body decoherence appears an attractive topic for future theoretical research -- much as the development of distillation \cite{Marshall-DI-2022} or error correction protocols to stabilize many-body coherences in actual experiments.

\begin{acknowledgements}
We thank Jonathan Brugger, Eric Brunner, Christian Haen, and Philipp Preiss for fruitful discussions. We are also thankful to Gabriel Dufour for fruitful discussions and comments at an early stage of the manuscript. C.D. acknowledges the Georg H. Endress Foundation for support and the Freiburg Institute for Advanced Studies for a FRIAS Junior Fellowship.
\end{acknowledgements}


%

\newpage

\title{Supplemental Material: Distinguishability-induced many-body decoherence}
\date{\today}

\maketitle
\onecolumngrid

\setcounter{equation}{0}
\renewcommand{\theequation}{S.\arabic{equation}}


\section{Simplified expressions of $\mathcal{W}_\mathrm{C}$}
In the following we provide a simplified expression for $\mathcal{W}_\mathrm{C}$ from Eq.~\eqref{eq:WC} of the main text, assuming many-body internal states of the form $\rho=\rho_\mathrm{1p} \otimes \cdots \otimes \rho_\mathrm{1p}$. To this end, we plug the matrix elements from Eq.~\eqref{eq:rhoEElements} of the main text into the definition of the many-body coherence from Eq.~\eqref{eq:WC} and use the cyclic property of the trace, 
\begin{align}
\begin{split}
\mathcal{W}_\mathrm{C}&=\frac{1}{N!(N!-1)}\sum_{\substack{\pi,\pi'\in\sg{N} \\ \pi\neq\pi'}} \abs{\tr{\Pi_\pi \rho \Pi^\dagger_{\pi'}} }\\
&=\frac{1}{N!-1} \left( \frac{1}{N!}\sum_{\pi,\pi'\in\sg{N} } \abs{\tr{\Pi^\dagger_{\pi'}\Pi_\pi \rho } } -1\right).
\end{split}
\end{align}
Since $\sg{N}$ forms a group, the summation can be reduced to
\begin{align}
\mathcal{W}_\mathrm{C}&=\frac{1}{N!-1} \left( \sum_{\pi\in\sg{N} } \abs{\tr{\Pi_\pi \rho } } -1\right). \label{eq:WCsimple}
\end{align}
As detailed in the main text, we now assume that all particles are in the same internal single-particle state $\rho_\mathrm{1p}$ such that $\rho=\rho_\mathrm{1p} \otimes \cdots \otimes \rho_\mathrm{1p}$. Following our considerations from the main text, we assume that $\rho_\mathrm{1p}$ has a discrete spectrum with $m$ eigenvalues $\lambda_j \geq 0$ and corresponding eigenvectors $\ket{j}$. Hence, its eigendecomposition reads
\begin{align}
\rho_\mathrm{1p}=\sum_{j=1}^m \lambda_j \ketbra{j}{j}
\end{align}
and the many-body internal state $\rho$ becomes
\begin{align}
\rho=\sum_{\vec{\mathcal{I}} \in \{1,\dots,m\}^N} \lambda_{\vec{\mathcal{I}}}\ \ketbra{\vec{\mathcal{I}}}{\vec{\mathcal{I}}}, \label{eq:rhoEigen}
\end{align}
where $\vec{\mathcal{I}}=(\mathcal{I}_1,\dots,\mathcal{I}_N)$ is the internal assignment list with $N$ elements $\mathcal{I}_\alpha \in \{1,\dots,m\}$, $\ket{\vec{\mathcal{I}}}=\ket{\mathcal{I}_1}\otimes \cdots \otimes \ket{\mathcal{I}_N}$, and $\lambda_{\vec{\mathcal{I}}}=\prod_{\alpha=1}^N \lambda_{\mathcal{I}_\alpha}$. Using this, the traces in Eq.~\eqref{eq:WCsimple} become
\begin{align}
\begin{split}
\tr{\Pi_\pi \rho }&=\sum_{\vec{\mathcal{I}} \in \{1,\dots,m\}^N} \lambda_{\vec{\mathcal{I}}} \ \tr{ \Pi_\pi \ketbra{\vec{\mathcal{I}}}{\vec{\mathcal{I}}} }\\
&=\sum_{\vec{\mathcal{I}} \in \{1,\dots,m\}^N} \lambda_{\vec{\mathcal{I}}} \ \bracket{\vec{\mathcal{I}}}{\vec{\mathcal{I}}_\pi}, 
\end{split}\label{eq:trpirho}
\end{align}
where $\Pi_\pi \ket{\vec{\mathcal{I}}}=\ket{\vec{\mathcal{I}}_\pi}=\ket{\mathcal{I}_{\pi(1)}}\otimes \dots \otimes \ket{\mathcal{I}_{\pi(N)}}$. Thus, since $\bracket{\vec{\mathcal{I}}}{\vec{\mathcal{I}}_\pi} =\prod_{\alpha=1}^N \delta_{\mathcal{I}_\alpha,\mathcal{I}_{\pi(\alpha)}} \geq 0$ and $\lambda_{\vec{\mathcal{I}}} \geq 0$, we have $\tr{\Pi_\pi \rho }\geq 0$. Accordingly, we can drop the modulus in~\eqref{eq:WCsimple}, yielding the simplified expression
\begin{align}\label{eq:WCsimplederiv}
\mathcal{W}_\mathrm{C}&=\frac{1}{N!-1} \left( \sum_{\pi\in\sg{N} } \tr{\Pi_\pi \rho } -1\right).
\end{align}
With the help of the projector $\Pi_\mathrm{S}=1/N! \sum_{\pi \in \mathrm{S}_N} \Pi_\pi$ onto the symmetric $N$-particle subspace, this can further be written as
\begin{align}\label{eq:WCproequ}
\mathcal{W}_\mathrm{C}&=\frac{N!}{N!-1}  \tr{\Pi_\mathrm{S} \rho } -\frac{1}{N!-1},
\end{align}
which, in the limit $N!\gg1$, simplifies to 
\begin{align}\label{eq:WCprojector}
\mathcal{W}_\mathrm{C}\approx\tr{\Pi_\mathrm{S} \rho }.
\end{align}

\section{Relation between $\mathcal{W}_\mathrm{C}$ and the expectation values $\tr{\Pi_\mathrm{S(A)} \rho_\mathrm{E}}$ and $\tr{\Pi_\mathrm{S} \rho}$}
\label{sec:eigenvalues}
To calculate the expectation value of the projector $\Pi_\mathrm{S(A)}=1/N! \sum_{\tau \in \mathrm{S}_N} (-1)^\tau _\mathrm{B(F)} \Pi_\tau $ onto the (anti)symmetric $N$-particle subspace with respect to the particles' external degrees of freedom, we use the matrix elements of $\rho_\mathrm{E}$ as provided in Eq.~\eqref{eq:rhoEElements} of the main text, 
\begin{align}
\begin{split}
\tr{\Pi_\mathrm{S(A)}\rho_\mathrm{E}}&=\frac{1}{N!^2} \sum_{\tau,\pi,\pi' \in \mathrm{S}_N} (-1)^{\tau\pi\pi'}_\mathrm{B(F)} \tr{ \Pi_\pi \rho \Pi_{\pi'}^\dagger} \bra{\vec{E}_{\pi'}} \Pi_\tau \ket{\vec{E}_\pi} \\
&=\frac{1}{N!^2} \sum_{\tau,\pi,\pi' \in \mathrm{S}_N} (-1)^{\tau\pi\pi'}_\mathrm{B(F)} \tr{ \Pi_{\pi(\pi')^{-1}} \rho} \bracket{\vec{E}}{\vec{E}_{\pi\tau (\pi')^{-1}}} \\
&=\frac{1}{N!^2} \sum_{\pi,\pi' \in \mathrm{S}_N}  \tr{ \Pi_{\pi(\pi')^{-1}} \rho} \\
&=\frac{1}{N!} \sum_{\pi\in \mathrm{S}_N}  \tr{ \Pi_\pi \rho} \\
&=\tr{\Pi_\mathrm{S} \rho}.
\end{split}
\end{align}
That is, we just showed that $\tr{\Pi_\mathrm{S(A)}\rho_\mathrm{E}}=\tr{\Pi_\mathrm{S} \rho}$. Using this in Eq.~\eqref{eq:WCprojector} finally yields
\begin{align}
\mathcal{W}_\mathrm{C}\approx\tr{\Pi_\mathrm{S(A)}}=\tr{\Pi_\mathrm{S} \rho},
\end{align}
as stated in the main text.

\section{Proof of Eq.~(\ref{eq:WCeigen})}\label{sec:spectrum}
In the following we prove Eq.~\eqref{eq:WCeigen} of the main text. We start wit plugging Eq.~\eqref{eq:trpirho} into Eq.~\eqref{eq:WCsimplederiv},
\begin{align}
\mathcal{W}_\mathrm{C}&=\frac{1}{N!-1} \left(  \sum_{\vec{\mathcal{I}} \in \{1,\dots,m\}^N} \lambda_{\vec{\mathcal{I}}} \sum_{\pi\in\sg{N} } \bracket{\vec{\mathcal{I}}}{\vec{\mathcal{I}}_\pi}  -1\right).\label{eq:WCsumII}
\end{align}
Now, let us introduce the internal occupation list $\vec{J}=(J_1,\dots,J_m)$, with $J_j$ the number of particles in the internal state $\ket{j}$ such that $\sum_{j=1}^m J_j=N$. Note that an internal occupation $\vec{J}$ can give rise to several assignment lists $\vec{\mathcal{I}}$ [see below Eq.~\eqref{eq:rhoEigen}], which differ by permutations of their elements. In particular, for the internal occupation $\vec{J}$, let $\vec{I}$ be the corresponding assignment list whose elements are listed in ascending order. Using the notation $\ket{\vec{I}_\pi}=\ket{I_{\pi(1)}} \otimes \cdots \otimes \ket{I_{\pi(N)}}$ for $\pi\in\sg{N}$, we see that $\ket{\vec{I}_\xi}=\ket{\vec{I}}$ if and only if $\xi \in \sg{\vec{J}} = \sg{J_1}\otimes\cdots\otimes \sg{J_m}$, which is a Young subgroup of $\sg{N}$. Hence, for $\pi\in\sg{N}$ all permutations of the right coset $\sg{\vec{J}}\pi=\{\xi\pi|\xi\in\sg{\vec{J}}\}$ result in the same state, i.e., $\ket{\vec{I}_\pi}=\ket{\vec{I}_{\pi'}}$ for all $\pi'\in \sg{\vec{J}}\pi$. Therefore, let us construct the transversal $\Sigma(\vec{J})$ of the set of right cosets of $\sg{\vec{J}}$ in $\sg{N}$ containing one permutation of each distinct right coset such that $\bracket{\vec{I_\mu}}{\vec{I_\nu}}=\delta_{\mu,\nu}$ for $\mu,\nu \in \Sigma(\vec{J})$  \cite{Dittel-AI-2019}. Note that $\Sigma(\vec{J})$ has cardinality $J\equiv |\Sigma(\vec{J})| = N!/\prod_{j=1}^m J_j!$. Using this, we can rewrite the sum over all assignment lists $\vec{\mathcal{I}}$ in Eq.~\eqref{eq:WCsumII}, resulting in
\begin{align}
\mathcal{W}_\mathrm{C}&=\frac{1}{N!-1} \left(   \sum_{\vec{J}} \sum_{\mu \in \Sigma(\vec{J})} \lambda_{\vec{I}_\mu} \sum_{\pi\in\sg{N} } \bracket{\vec{I}_\mu}{\vec{I}_{\mu\pi}}  -1\right).
\end{align}
Using $\sum_{\pi\in\sg{N} } \bracket{\vec{I}_\mu}{\vec{I}_{\mu\pi}}=N!/J$, $\lambda_{\vec{I}_\mu}=\lambda_{\vec{I}} \equiv \lambda_{\vec{J}} $, with $ \lambda_{\vec{J}} =\prod_{j=1}^m \lambda_j^{J_j}$, this simplifies to 
\begin{align}
\begin{split}
\mathcal{W}_\mathrm{C}&=\frac{1}{N!-1} \left(   \sum_{\vec{J}} \sum_{\mu \in \Sigma(\vec{J})}  \frac{N!}{J}  \lambda_{\vec{J}} -1\right)\\
&=\frac{1}{N!-1} \left(   N! \sum_{\vec{J}} \lambda_{\vec{J}} -1\right)\\
&=\frac{N!}{N!-1} \sum_{\vec{J}} \lambda_{\vec{J}}  - \frac{1}{N!-1}.
\end{split}
\end{align}
Accordingly, with the help of Eq.~\eqref{eq:WCproequ} we find $\tr{\Pi_\mathrm{S} \rho}= \sum_{\vec{J}} \lambda_{\vec{J}}$ such that for $N!\gg 1 $, we get
\begin{align}\label{eq:WCEigapp}
\mathcal{W}_\mathrm{C}&\approx \sum_{\vec{J}} \lambda_{\vec{J}}.
\end{align}
Using $ \lambda_{\vec{J}} =\prod_{j=1}^m \lambda_j^{J_j}$ and $\sum_{j=1}^m J_j=N$, this can also be written as
\begin{align}\label{eq:WCEigappasmain}
\mathcal{W}_\mathrm{C}&\approx \sum_{J_1+J_2+\cdots+J_m=N} \lambda_1^{J_1}\lambda_2^{J_2} \cdots \lambda_m^{J_m},
\end{align}
which coincides with Eq.~\eqref{eq:WCeigen} of the main text.

\section{Proof of Eq.~(\ref{eq:WCThLimit})}\label{sec:thermlimit}
We start our proof of Eq.~\eqref{eq:WCThLimit} of the main text by considering $\mathcal{W}_\mathrm{C}$ from Eq.~\eqref{eq:WCeigen} of the main text [see also Eq.~\eqref{eq:WCEigappasmain}]. Without loss of generality we can assume that the eigenvalues of $\rho_\mathrm{1p}$ satisfy $\lambda_1 \leq \lambda_2 \leq \cdots \leq \lambda_m$. That is, $\lambda_\mathrm{max}=\lambda_m$ is the maximal eigenvalue of $\rho_\mathrm{1p}$. We further suppose that $\lambda_\mathrm{m}$ is $d$-fold degenerate, i.e., $\lambda_m=\lambda_{m-1}=\cdots=\lambda_{m+1-d}$. With this in mind, let us rewrite $\mathcal{W}_\mathrm{C}$ from Eq.~\eqref{eq:WCEigapp},
\begin{align}
\begin{split}
\mathcal{W}_\mathrm{C} &\approx \sum_{J_1+J_2+\cdots+J_m=N} \lambda_1^{J_1}\lambda_2^{J_2} \cdots \lambda_m^{J_m}\\
&= \sum_{J_1=0}^N \lambda_1^{J_1} \sum_{J_2=0}^N \lambda_2^{J_2} \cdots \sum_{J_{m-1}=0}^N \lambda_{m-1}^{J_{m-1}} \ \lambda_m^{N-J_1-J_2-\dots -J_{m-1}} \ \Theta(N-J_1-J_2-\dots-J_{m-1}) \\
&= \lambda_m^N \sum_{J_1=0}^N \left( \frac{\lambda_1}{\lambda_m} \right)^{J_1} \sum_{J_2=0}^N \left( \frac{\lambda_2}{\lambda_m} \right)^{J_2}\cdots \sum_{J_{m-1}=0}^N \left( \frac{\lambda_{m-1}}{\lambda_m}\right)^{J_{m-1}} \  \Theta(N-J_1-J_2-\dots-J_{m-1}),
\end{split}\label{eq:WCHev1}
\end{align}
where 
\begin{align}\label{eq:HSF}
\Theta(N-J_1-J_2-\dots-J_{m-1})=\begin{cases}  1\quad \text{for } J_1+J_2+\dots+ J_{m-1} \leq N \\
0\quad \text{otherwise}\end{cases}
\end{align}
is the Heaviside function. By the degeneracy of the maximum eigenvalue, $\lambda_m=\lambda_{m-1}=\cdots=\lambda_{m+1-d}$, this becomes 
\begin{align}
\mathcal{W}_\mathrm{C} &\approx \lambda_m^N \sum_{J_1=0}^N \left( \frac{\lambda_1}{\lambda_m} \right)^{J_1} \cdots \sum_{J_{m-d}=0}^N \left( \frac{\lambda_{m-d}}{\lambda_m}\right)^{J_{m-d}} \sum_{J_{m+1-d}=0}^N \cdots \sum_{J_{m-1}=0}^N \  \Theta(N-J_1-J_2-\dots-J_{m-1}).
\end{align}
Now note that in the limit $N\rightarrow\infty$, the Heaviside function~\eqref{eq:HSF} can be approximated by unity. Hence, in the limit $N\rightarrow\infty$ the many-body coherence is well approximated by
\begin{align}
\mathcal{W}_\mathrm{C} &\approx \lambda_m^N \sum_{J_1=0}^N \left( \frac{\lambda_1}{\lambda_m} \right)^{J_1} \cdots \sum_{J_{m-d}=0}^N \left( \frac{\lambda_{m-d}}{\lambda_m}\right)^{J_{m-d}} \sum_{J_{m+1-d}=0}^N \cdots \sum_{J_{m-1}=0}^N  \label{eq:WCthlimd}
\end{align}
Next, we use the geometric series 
\begin{align}
\sum_{J_j=0}^N \left(\frac{\lambda_j}{\lambda_m}\right)^{J_j}=\left(1-\frac{\lambda_j}{\lambda_m}\right)^{-1}- \left(\frac{\lambda_j}{\lambda_m}\right)^{N+1}\left(1-\frac{\lambda_j}{\lambda_m}\right)^{-1}
\end{align}
such that $\mathcal{W}_\mathrm{C}$ from Eq.~\eqref{eq:WCthlimd} becomes
\begin{align}
\mathcal{W}_\mathrm{C} &\approx (N+1)^{d-1} \ \lambda_m^N \ \prod_{j=1}^{m-d} \left[\left(1-\frac{\lambda_j}{\lambda_m}\right)^{-1} - \left(\frac{\lambda_j}{\lambda_m}\right)^{N+1}\left(1-\frac{\lambda_j}{\lambda_m}\right)^{-1}  \right].
\end{align}
After performing the product, we see that there are factors of the form $\lambda_m^N$, $\lambda_m^N (\lambda_j/\lambda_m)^{N+1}$, $\lambda_m^N (\lambda_j/\lambda_m)^{N+1} (\lambda_k/\lambda_m)^{N+1}$, etc. However, since $\lambda_j/\lambda_m<1$ for all $j=1,\dots,m-d$, in the limit $N\rightarrow \infty$ the dominant factors are those of the form $\lambda_m^N$. That is, the approximation of $\mathcal{W}_\mathrm{C}$ simplifies to
\begin{align}
\mathcal{W}_\mathrm{C} &\approx (N+1)^{d-1} \ \lambda_m^N \ \prod_{j=1}^{m-d} \left(1-\frac{\lambda_j}{\lambda_m}\right)^{-1}.\label{eq:WCthlimd2}
\end{align}
Note that one arrives at the same result faster by setting the upper limit of the first $m-d$ sums in~\eqref{eq:WCthlimd} to infinity. Equation~\eqref{eq:WCthlimd2} describes the behaviour of $\mathcal{W}_\mathrm{C}$ in the thermodynamic limit. In the case of a non-degenerate maximum eigenvalue $\lambda_m$, i.e., $d=1$, the approximation simplifies to
\begin{align}
\mathcal{W}_\mathrm{C} &\approx \lambda_m^N\ \prod_{j=1}^{m-1} \left(1-\frac{\lambda_j}{\lambda_m}\right)^{-1}
\end{align}
as stated in Eq.~\eqref{eq:WCThLimit} of the main text.

\section{Proof of Eq.~(\ref{eq:WCdecay})}\label{sec:faint}
In the following we prove Eq.~\eqref{eq:WCdecay} of the main text. As stated in the main text, in the case of small deviations from perfectly indistinguishable particles, $\rho_\mathrm{1p}$ can be written as $\rho_\mathrm{1p} =(1-\epsilon)\ketbra{\phi}{\phi} + \epsilon \tilde{\rho}_\mathrm{1p}$, with $\epsilon\ll 1/2$. Note that $\tilde{\rho}_\mathrm{1p}$ is Hermitian and has unit trace. Using this decomposition of $\rho_\mathrm{1p}$, the many-body internal state $\rho$ becomes
\begin{align}
\begin{split}
\rho&=\left[(1-\epsilon)\ketbra{\phi}{\phi} + \epsilon \tilde{\rho}_\mathrm{1p} \right] \otimes \cdots \otimes \left[(1-\epsilon)\ketbra{\phi}{\phi} + \epsilon \tilde{\rho}_\mathrm{1p} \right]\\
&=(1-\epsilon)^N \ \ketbra{\phi}{\phi}\otimes \dots \otimes \ketbra{\phi}{\phi} \\
&+(1-\epsilon)^{N-1} \epsilon \left( \tilde{\rho}_\mathrm{1p} \otimes \ketbra{\phi}{\phi} \otimes \dots \otimes \ketbra{\phi}{\phi} +\dots + \ketbra{\phi}{\phi} \otimes \dots \otimes \ketbra{\phi}{\phi} \otimes \tilde{\rho}_\mathrm{1p}\right)\\
&+(1-\epsilon)^{N-2} \epsilon^2 \left( \tilde{\rho}_\mathrm{1p} \otimes \tilde{\rho}_\mathrm{1p} \otimes \ketbra{\phi}{\phi} \otimes \dots \otimes \ketbra{\phi}{\phi} +\dots + \ketbra{\phi}{\phi} \otimes \dots \otimes \ketbra{\phi}{\phi} \otimes \tilde{\rho}_\mathrm{1p}\otimes \tilde{\rho}_\mathrm{1p}\right)\\
&+\dots\label{eq:rho-app}
\end{split}
\end{align}
Now note that for $\epsilon \ll 1/2$, we have $(1-\epsilon)^{N-j} \epsilon^j \gg (1-\epsilon)^{N-j-1} \epsilon^{j+1}$. Accordingly, the first term of the sum in~\eqref{eq:rho-app} dominates. Therefore we can approximate $\mathcal{W}_\mathrm{C}$ from Eq.~\eqref{eq:WCsimplederiv} as
\begin{align}
\begin{split}
\mathcal{W}_\mathrm{C}&\approx\frac{1}{N!-1} \left( (1-\epsilon)^N \sum_{\pi\in\sg{N} } \tr{\Pi_\pi \ketbra{\phi}{\phi}\otimes \dots \otimes \ketbra{\phi}{\phi} } -1\right)\\
&= \frac{N!}{N!-1} (1-\epsilon)^N -\frac{1}{N!-1}.
\end{split}
\end{align}
In the limit  $N!\gg 1$ this simplifies to
\begin{align}\label{eq:WClimit}
\mathcal{W}_\mathrm{C}\approx (1-\epsilon)^N,
\end{align}
 as stated in Eq.~\eqref{eq:WCdecay} in the main text. 

\section{Lower order coherences}\label{sec:lowerorder}

Let us consider the reduced external state [see Eq.~\eqref{eq:rhoEElements} in the main text]
\begin{align}\label{eq:rhoEapp}
\rho_\mathrm{E}= \sum_{\pi,\pi' \in \sg{N}} [\rho_\mathrm{E} ]_{\pi,\pi'} \ketbra{\vec{E}_\pi}{\vec{E}_{\pi'}}
\end{align}
with 
\begin{align}\label{eq:rhoEpipiapp}
[\rho_\mathrm{E} ]_{\pi,\pi'}=(-1)_\mathrm{B(F)}^{\pi\pi'} \frac{1}{N!} \tr{\Pi_\pi \rho \Pi^\dagger_{\pi'}}.
\end{align}
By tracing out a particle, we obtain the reduced external $N-1$-particle state
\begin{align}
\begin{split}
\rho_\mathrm{E}^{(N-1)}&= \sum_{\pi,\pi' \in \sg{N}} [\rho_\mathrm{E} ]_{\pi,\pi'} \bracket{E_{\pi'(N)}}{E_{\pi(N)}}\ \ketbra{\vec{E}^{(N-1)}_\pi}{\vec{E}^{(N-1)}_{\pi'}}\\
&=\sum_{\substack{\pi,\pi' \in \sg{N} \\ \pi(N)=\pi'(N)}} [\rho_\mathrm{E} ]_{\pi,\pi'}\  \ketbra{\vec{E}^{(N-1)}_\pi}{\vec{E}^{(N-1)}_{\pi'}},
\end{split}
\end{align}
where $\ket{\vec{E}^{(N-1)}_{\pi}} = \ket{E_{\pi(1)}} \otimes \dots \otimes \ket{E_{\pi(N-1)}}$. With the help of the Young subgroup $\mathrm{S}_{N-1;\alpha}=\mathrm{S}_{\{1,\dots,N\}\setminus \{\alpha\}}\otimes \mathrm{S}_{\{\alpha\}}$, this can be written as
\begin{align}\label{eq:apprhoN1}
\begin{split}
\rho_\mathrm{E}^{(N-1)}&=\sum_{\alpha=1}^N \sum_{\pi,\pi' \in \mathrm{S}_{N-1;\alpha} } [\rho_\mathrm{E} ]_{\pi,\pi'}\  \ketbra{\vec{E}^{(N-1)}_\pi}{\vec{E}^{(N-1)}_{\pi'}} \\
&=\frac{1}{N}\sum_{\alpha=1}^N \sum_{\pi,\pi' \in \mathrm{S}_{N-1;\alpha} } [\rho_\mathrm{E}^{(N-1)} ]_{\pi,\pi'}\  \ketbra{\vec{E}^{(N-1)}_\pi}{\vec{E}^{(N-1)}_{\pi'}} \\
&=\frac{1}{N}\sum_{\alpha=1}^N \rho_\mathrm{E}^{(N-1;\alpha)},
\end{split}
\end{align}
where $[\rho_\mathrm{E}^{(N-1)}]_{\pi,\pi'}=N [\rho_\mathrm{E}]_{\pi,\pi'}$, and 
\begin{align}
\rho_\mathrm{E}^{(N-1;\alpha)}=\sum_{\pi,\pi' \in \mathrm{S}_{N-1;\alpha} } [\rho_\mathrm{E}^{(N-1)} ]_{\pi,\pi'}\  \ketbra{\vec{E}^{(N-1)}_\pi}{\vec{E}^{(N-1)}_{\pi'}}.
\end{align}
We recognize that $\rho_\mathrm{E}^{(N-1;\alpha)}$ corresponds to the reduced external state with the $\alpha$th particle excluded. Since we consider the internal product state $\rho=\rho_\mathrm{1p}\otimes \dots \otimes \rho_\mathrm{1p}$ in Eq.~\eqref{eq:rhoEpipiapp}, for different $\alpha$, the states $\rho_\mathrm{E}^{(N-1;\alpha)}$ only differ by the labeling of the external states, and, thus, must have equal many-body coherences. Accordingly, by Eq.~\eqref{eq:apprhoN1}, and the linearity of $\mathcal{W}_\mathrm{C}^{(N-1)}$ with respect to $\rho_\mathrm{E}^{(N-1)}$, the states $\rho_\mathrm{E}^{(N-1)}$ and $\rho_\mathrm{E}^{(N-1;N)}$ have equal many-body coherences. Thus, since 
\begin{align}
\rho_\mathrm{E}^{(N-1;N)}=\sum_{\pi,\pi' \in \mathrm{S}_{N-1} } [\rho_\mathrm{E}^{(N-1)} ]_{\pi,\pi'}\  \ketbra{\vec{E}^{(N-1)}_\pi}{\vec{E}^{(N-1)}_{\pi'}}
\end{align}
coincides with the reduced external state $\rho_\mathrm{E}$ from Eq.~\eqref{eq:rhoEapp} with $N-1$ instead of $N$ particles, we can conclude that the normalized many-body coherence $\mathcal{W}_\mathrm{C}^{(N-1)}$ of the reduced $N-1$-particle state $\rho_\mathrm{E}^{(N-1)}$ coincides with the normalized many-body coherence $\mathcal{W}_\mathrm{C}$ [see Eq.~\eqref{eq:WC} in the main text] in the case of $N-1$ instead of $N$ particles. Similarly, by tracing out further particles, the same reasoning lets us conclude that the many-body coherence $\mathcal{W}_\mathrm{C}^{(k)}$ of the reduced $k$-particle state $\rho_\mathrm{E}^{(k)}=\trpp{N-k}{\rho_\mathrm{E}}$ coincides with $\mathcal{W}_\mathrm{C}$ in the case of $k$ instead of $N$ particles.

The same conclusion can be drawn faster by considering the expression of $\mathcal{W}_\mathrm{C}$ from Eq.~\eqref{eq:WCproequ}: Since each term of the reduced $k$-particle state $\rho_\mathrm{E}^{(k)}$ is associated with the unsymmetrized internal state $\rho^{(k)}=\rho_\mathrm{1p}^{\otimes k}$, by Eq.~\eqref{eq:WCproequ}, we must have
\begin{align}
\mathcal{W}_\mathrm{C}^{(k)}&=\frac{k!}{k!-1}  \tr{\Pi_\mathrm{S}^{(k)} \rho^{(k)} } -\frac{1}{k!-1},
\end{align}
with $\Pi_\mathrm{S}^{(k)} = 1/k! \sum_{\pi\in\mathrm{S}_k} \Pi_\pi$. 

\section{Atoms in the small temperature limit}\label{sec:faintAtom}
In the small temperature limit $\kb T \ll \Delta E/\ln(2)$ the atoms are with hight probability ($e^{-\beta E_1} Z(\beta)^{-1}$) in the internal ground state $\ket{1}$. Hence, by rewriting the single-particle internal state $\rho_\mathrm{1p}= \sum_{j=1}^{m} e^{-\beta E_j}Z(\beta)^{-1} \ketbra{j}{j}$ as
\begin{align}
\begin{split}
\rho_\mathrm{1p}&=\frac{e^{-\beta E_1}}{Z(\beta)} \ketbra{1}{1}+ \sum_{j=2}^m \frac{e^{-\beta E_j}}{Z(\beta)} \ketbra{j}{j}\\
&=(1-\epsilon) \ketbra{1}{1}+\epsilon\ \tilde{\rho}_\mathrm{1p},
\end{split}
\end{align} 
we can identify 
\begin{align}
1-\epsilon=\frac{e^{-\beta E_1}}{Z(\beta)} \approx \frac{e^{-\beta E_1}}{e^{-\beta E_1}+e^{-\beta E_2}} = \frac{1}{1+e^{-\beta \Delta E}} \approx 1- e^{-\beta \Delta E},
\end{align}
i.e., we have $\epsilon \approx e^{-\beta \Delta E}$. For particle numbers $N! \gg 1$ we can then apply Eq.~\eqref{eq:WCdecay} [see also Eq.~\eqref{eq:WClimit}], resulting in 
\begin{align}
\mathcal{W}_\mathrm{C} \approx (1- e^{-\beta \Delta E})^N
\end{align}
as stated in the main text.

\section{Faint distinguishabilities of photons with random arrival times}\label{sec:faintPhoton}
Let us consider the single-particle internal state $\rho_\mathrm{1p}=\int \mathrm{d}t\ P(t) \ketbra{t}{t}$ for any probability distribution $P(t)$, i.e., $P(t)$ is not necessarily a normal distribution. Without loss of generality we suppose that $\braket{t}=\int \mathrm{d}t\ t P(t) =0$ such that, in the case of faint distingiushabilities $\sigma\Delta \ll 1/\sqrt{2}$, the photons arrive at time $t\approx \braket{t}=0$ with high probability. Hence, we can write the single-particle internal state as
\begin{align}
\begin{split}
\rho_\mathrm{1p}&= \bra{0}\rho_\mathrm{1p}\ket{0} \ketbra{0}{0} + (1-\bra{0}\rho_\mathrm{1p}\ket{0}) \tilde{\rho}_\mathrm{1p}\\
&=(1-\epsilon) \ketbra{0}{0} + \epsilon\ \tilde{\rho}_\mathrm{1p},
\end{split}
\end{align}
and identify $1-\epsilon =\bra{0}\rho_\mathrm{1p}\ket{0} $. Calculating the expectation value yields
\begin{align}
\begin{split}
\bra{0}\rho_\mathrm{1p}\ket{0} &=\int_{-\infty}^\infty \mathrm{d}t\ P(t) \abs{\bracket{0}{t}}^2\\
&=\int_{-\infty}^\infty \mathrm{d}t\ P(t) e^{-\Delta^2 t^2},\label{eq:IntPt}
\end{split}
\end{align}
where we used $\ket{t}=(2 \pi \Delta^2)^{-1/4} \int_{-\infty}^\infty \mathrm{d}\omega \ e^{\im \omega t} e^{-\frac{(\omega-\Omega)^2}{4\Delta^2}} \ket{\omega}$ to obtain the overlap $\abs{\bracket{0}{t}}^2=e^{-\Delta^2 t^2}$. Since $P(t)$ in~\eqref{eq:IntPt} must be small for times $|t|>1/\Delta$ [recall that we consider faint disinguishabilities $\sigma\Delta \ll 1/\sqrt{2}$, i.e., $\sigma \ll 1/\sqrt{2}\Delta$], we can expand the exponential function in~\eqref{eq:IntPt}, resulting in
\begin{align}
\begin{split}
\bra{0}\rho_\mathrm{1p}\ket{0}  &\approx \int_{-\infty}^\infty \mathrm{d}t\ P(t) \left( 1-\Delta^2 t^2\right)\\
&=1-\Delta^2 \sigma^2,
\end{split}
\end{align}
where we used $\braket{t^2}=\sigma^2$, since $\braket{t}=0$. Hence, we find $1-\epsilon \approx 1-\Delta^2 \sigma^2$ (compare to the average mutual fidelity in \cite{SC-Shchesnovich-2014}). By Eq.~\eqref{eq:WCdecay} [see also Eq.~\eqref{eq:WClimit}], for $N!\gg 1$, we then have $\mathcal{W}_\mathrm{C}\approx (1-\Delta^2 \sigma^2)^N$ as provided in the main text.

\end{document}